\documentclass[]{llncs}

\usepackage[T1]{fontenc}
\usepackage{cite}

\usepackage{amsmath, amssymb, amsfonts}
\usepackage[cmintegrals]{newtxmath}
\usepackage{bm}
\usepackage{array}

\usepackage{graphicx}
\usepackage{floatrow}
\usepackage{textcomp}
\usepackage[english]{babel}
\usepackage{blindtext}
\usepackage[utf8]{inputenc}
\usepackage{xcolor}
\usepackage{colortbl}
\usepackage[linesnumbered,algo2e,ruled]{algorithm2e}

\usepackage{tabularx}
\usepackage{algorithm}
\usepackage{listings}
\usepackage[thinlines]{easytable}
\usepackage{algpseudocode}
\usepackage{svg}
\usepackage{float}
\usepackage{pifont}

\usepackage{tablefootnote}
\usepackage{longtable}
\usepackage{multirow}
\usepackage{booktabs}

\usepackage[export]{adjustbox}[2011/08/13]

\usepackage[hyphens]{url}
\usepackage[hidelinks, bookmarks=false]{hyperref}
\usepackage{siunitx}
\usepackage{booktabs,subcaption,dcolumn}
\usepackage{soul}

\usepackage{mathtools, nccmath}
\usepackage{comment}

\usepackage[linesnumbered,algo2e,ruled]{algorithm2e}

\usepackage{tikz}

\newcommand*\halfcirc[1][1ex]{%
  \begin{tikzpicture}
  \draw[fill] (0,0)-- (90:#1) arc (90:270:#1) -- cycle ;
  \draw (0,0) circle (#1);
  \end{tikzpicture}}
\newcommand*\fullcirc[1][1ex]{\tikz\fill (0,0) circle (#1);} 

\newcommand\revision[1]{%
  \bgroup
  \hskip0pt\color{blue!80!black}%
  #1%
  \egroup
}

\newcommand{\xmark}{{\ding{55}}}

\begin{document}

\title{Problem space structural adversarial attacks for Network Intrusion Detection Systems based on Graph Neural Networks}

\newcommand*{\CtoXB}{$\text{\textbf{C2x}}_\mathbf{\mathcal{B}}$}
\newcommand*{\CtoXM}{$\text{\textbf{C2x}}_\mathbf{\mathcal{M}}$}

\author{
    Andrea Venturi\inst{1}* \and
    Dario Stabili\inst{2} \and
    Mirco Marchetti\inst{1}
}

\institute{
    University of Modena and Reggio Emilia\and \stepc University of Bologna
}

\maketitle

\begin{abstract}
Machine Learning (ML) algorithms have become increasingly popular for supporting Network Intrusion Detection Systems (NIDS). Nevertheless, extensive research has shown their vulnerability to adversarial attacks, which involve subtle perturbations to the inputs of the models aimed at compromising their performance. Recent proposals have effectively leveraged Graph Neural Networks (GNN) to produce predictions based also on the structural patterns exhibited by intrusions to enhance the detection robustness. However, the adoption of GNN-based NIDS introduces new types of risks. In this paper, we propose the first formalization of adversarial attacks specifically tailored for GNN in network intrusion detection. Moreover, we outline and model the problem space constraints that attackers need to consider to carry out feasible structural attacks in real-world scenarios. As a final contribution, we conduct an extensive experimental campaign in which we launch the proposed attacks against state-of-the-art GNN-based NIDS. Our findings demonstrate the increased robustness of the models against classical feature-based adversarial attacks, while highlighting their susceptibility to structure-based attacks.
\end{abstract}

\section{Introduction}

Machine Learning (ML) solutions are increasingly being integrated into modern Network Intrusion Detection Systems (NIDS). While enhancing the detection performance, ML algorithms are also vulnerable to adversarial attacks, which involve subtle perturbations to input samples aimed at misleading the target detector into producing an incorrect classification~\cite{biggio2014security}. Despite considerable research efforts to devise effective countermeasures, adversarial attacks remain an open problem, with no universally effective solution available for mitigating all types of attacks or applicable to all ML models~\cite{apruzzese2023role}.

With the rise of Deep Learning (DL), a new generation of DL-based NIDS has emerged. Among these, NIDS based on Graph Neural Networks (GNN) appear to offer a promising detection approach~\cite{bilot2023graph}. The idea behind the employment of these models is to overcome the limitations of traditional ML-based NIDS, by providing robust classifications leveraging not only on the patterns offered by cyberattacks at the feature level, but also on the topological structures they exhibit in the network. GNN-based NIDS process network traffic in a graph representation, providing classifications for each node or edge in the considered graph. 

Previous works for GNN-based NIDS have primarily demonstrated their effectiveness in non-adversarial settings~\cite{lo2022graphsage, venturi2023arganids}, or have limited their scope to \textit{feature}-based attacks, which involve perturbations that are often unfeasible in practice~\cite{zhou2021hierarchical, pujol2022unveiling, wang2022threatrace}. In contrast, in this paper, we are the first to introduce and design novel \textit{structural} adversarial attacks tailored for GNN-based NIDS. Instead of modifying the features of network flows as in traditional adversarial attacks, we consider realistic adversaries that are able to modify the graph structure submitted to the GNN in order to evade detection of their malicious communications. The aim is to affect the final predictions of the model by injecting new nodes or new edges into the graph, operating directly at the network traffic level. 

Launching these attacks in the network domain presents unique challenges compared to other contexts~\cite{pierazzi2020intriguing, apruzzese2022modeling}.
Therefore, as a further contribution of this paper, we formally identify and discuss the constraints that restrict the scope of an attacker's strategy when implementing structural adversarial attacks in real-world scenarios. Moreover, we present four novel structural attack variants that verify these constraints, and detail practical strategies for their execution. 

We evaluate our proposals in a vast experimental campaign, employing two inductive GNN-based NIDS, achieving state-of-the-art performance on two publicly available datasets. We explore multiple adversarial scenarios that comprise both traditional feature-based attacks that have been proven effective against conventional ML-based NIDS, and our innovative structural attacks that have never been considered before. The results demonstrate the enhanced robustness provided by the GNN-based NIDS to feature-based attacks with respect to detectors based on classical ML approaches. However, our findings also expose the extreme vulnerability of these models to structural perturbations. Our experiments show that, in most cases, it is sufficient for an attacker to set up a single new communication from the compromised nodes to arbitrary network targets, to drastically reduce the detection performance to inadequate levels. We will release the entire source code repository to reproduce all our evaluations.

The remainder of this paper is organized as follows. Section~\ref{sec:background} provides background details on GNN-based NIDS and on adversarial attacks. Section~\ref{sec:taxonomy} introduces adversarial attacks against GNN-based NIDS and outlines the problem-space constraints crucial for the development of structural attack in practice, presenting the four proposed attack variants. Section~\ref{sec:testbed} describes our experimental setup. Section~\ref{sec:results} reports and discusses the results. Section~\ref{sec:related} compares our work with relevant related research. Section~\ref{sec:conclusion} concludes the paper with some final remarks. 


\section{Background}
\label{sec:background}

\subsection{GNN-based NIDS}
\label{sec:gnn_nids}

GNN are innovative models that are able to handle graph-structured data. As network traffic can be naturally represented in a graph format, the idea is that GNN can capture both feature and topological patterns of cyberattacks to enhance detection performance.

There are three main phases for developing a GNN-based NIDS. The first phase involves the construction of the graph from a set of labeled \textit{netflows}\footnote{Netflows consist in tabular data structures commonly employed in network intrusion detection for resuming the communications between two endpoints with metadata and features.}~\cite{cisco2021netflow}. 
A graph $G$ can be generally defined as $G = (V, E)$, where $V$ represents a set of vertices or \textit{nodes}, and $E$ represents the \textit{edges} connecting the nodes. 
There are two principal graph structures used for network intrusion detection in literature. The most straightforward way to generate a suitable graph is by considering each netflow as an edge of the graph, connecting two nodes that correspond to the endpoint indicated on the netflow itself (i.e., \textit{flow graph}). 
In this case, the network intrusion detection task is equivalent to performing classification of the edges of the graph.

Although this might be the most natural approach to translating network data in a graph, most GNN are unable to consider features associated with the edges. For this reason, a dual approach (i.e., \textit{line graph}) has emerged~\cite{chang2021graph, venturi2023arganids}. The idea is that each edge of the original flow graph is transformed into a node, and two nodes in the line graph are connected if the corresponding edges in the flow graph share an endpoint in common. This process - called linearization - allows to associate the features of the netflows to each node of the graph, rather than to the edges, without losing topological information. With line graphs, the intrusion detection task is translated to a node classification problem. 

For the sake of this paper, it is important to remark that it is possible to generate the graph only \textit{after} that a set of netflows has been produced and collected. Hence, the two steps can be considered as independent and separate, as demonstrated by previous work~\cite{venturi2023practical}. 

After having obtained a suitable graph, the second phase involves the actual GNN for node embedding. This concept involves generating a node embedding through dimensionality reduction techniques, aiming to encapsulate high-dimensional information about a node's \textit{neighborhood} within a compact and dense vector~\cite{kipf2016semi}. The idea behind the adversarial attacks that we propose in this paper is to exploit this property of GNN, introducing alterations to the neighborhood of a node that causes a misclassification. Two main strategies exist for producing such embeddings. Most approaches follow a \textit{transductive} strategy, in which the models learn to optimize the embeddings in a single and fixed graph. However, these methods are unable to generalize to unseen data, as required by many real-world applications (e.g., network intrusion detection). On the other hand, \textit{inductive} GNN can generate embeddings even for unseen nodes or entirely new subgraphs~\cite{hamilton2017inductive}. This allows to simultaneously learn the topological structure of the node's neighborhood and the distribution of the node features in the neighborhood even if operating on dynamic graphs. Due to the intrinsic dynamic nature of network traffic, inductive GNN are more suitable for intrusion detection. For these reasons, we will only consider inductive GNN in the evaluation campaign of this paper.

The embeddings produced in the second phase can thus be associated with the respective labels and fed to machine learning systems to perform the classification (third phase). The idea is that operating directly on this more expressive data structure enhances the detection capabilities and robustness.

\subsection{Adversarial attacks}
\label{sec:adversarial_attacks}
This paper delves into the growing concern around the susceptibility of ML and DL algorithms to adversarial attacks, which consist of subtle perturbations operated on input data, leading the model to predict incorrect values or to learn wrong decision functions. 
Building upon previous works~\cite{biggio2014security, biggio2018wild}, we introduce a comprehensive taxonomy that classifies adversarial attacks based on four crucial dimensions: attacker's \textit{goal}, \textit{knowledge}, \textit{capabilities} and \textit{strategy}. 

The attacker's goal is expressed in terms of which security property of the machine learning model they are trying to violate, which are confidentiality, integrity and availability. Among those, attacks against integrity represent the most popular choice in research, and they are also the category we consider. Their aim is to evade detection without compromising the normal operations of the model~\cite{apruzzese2020deep}. 

The level of knowledge that an attacker has about the ML system is another important aspect to consider. White-box attacks operate under the assumption that the adversary possesses complete knowledge of the target system~\cite{sadeghzadeh2021adversarial}, often rendering them highly effective yet impractical. Conversely, black-box attacks treat the ML model as an unknown entity capable of receiving inputs and providing outputs (i.e., predictions). These are more adaptable to real-world scenarios, but require crafting adversarial samples with minimal information~\cite{usama2019black}. The middle ground between these extremes is gray-box attacks, wherein the attacker has partial knowledge about the model or its training data~\cite{apruzzese2020deep}.

The attacker's capability distinguishes between two primary types of actions. Poisoning attacks involve manipulating training data to induce the model to learn incorrect decision boundaries through the introduction of perturbed samples~\cite{papadopoulos2021launching}. Conversely, evasion attacks modify test samples to deceive the classifier into producing incorrect predictions~\cite{apruzzese2018evading}.

The last category involves the strategy followed by the attacker to craft the adversarial examples. 
The most general strategy involves changes to the feature values of a sample. Notably, it is important to differentiate between attacks exclusively operating in feature space, irrespective of the underlying problem formulation, and those necessitating consideration of the problem constraints within the application domain, resulting in the creation of real evasive objects.
\section{Adversarial attacks for GNN-based NIDS}
\label{sec:taxonomy}
In this section, we present the first formal analysis of adversarial attacks possible against GNN-based NIDS at test time. While previous works consider at most attacks perturbing the features of netflow data (\textit{feature attacks}), we propose a novel set of attacks involving perturbation to the graph structure itself (\textit{structural attacks}). In continuity with related work, we assume GNN-based NIDS operating on \textit{flow graphs}.
We recall that in flow graphs, the nodes denote the endpoints of the network (e.g., hosts), and the edges store the features representing the communications between these endpoints (e.g., netflows). We remark, however, that our notation can directly translate to line graph representations.

\subsection{Threat model}
\label{sec:threat_model}

\begin{figure}[tbp]
    \centering
    \includegraphics[width=\textwidth]{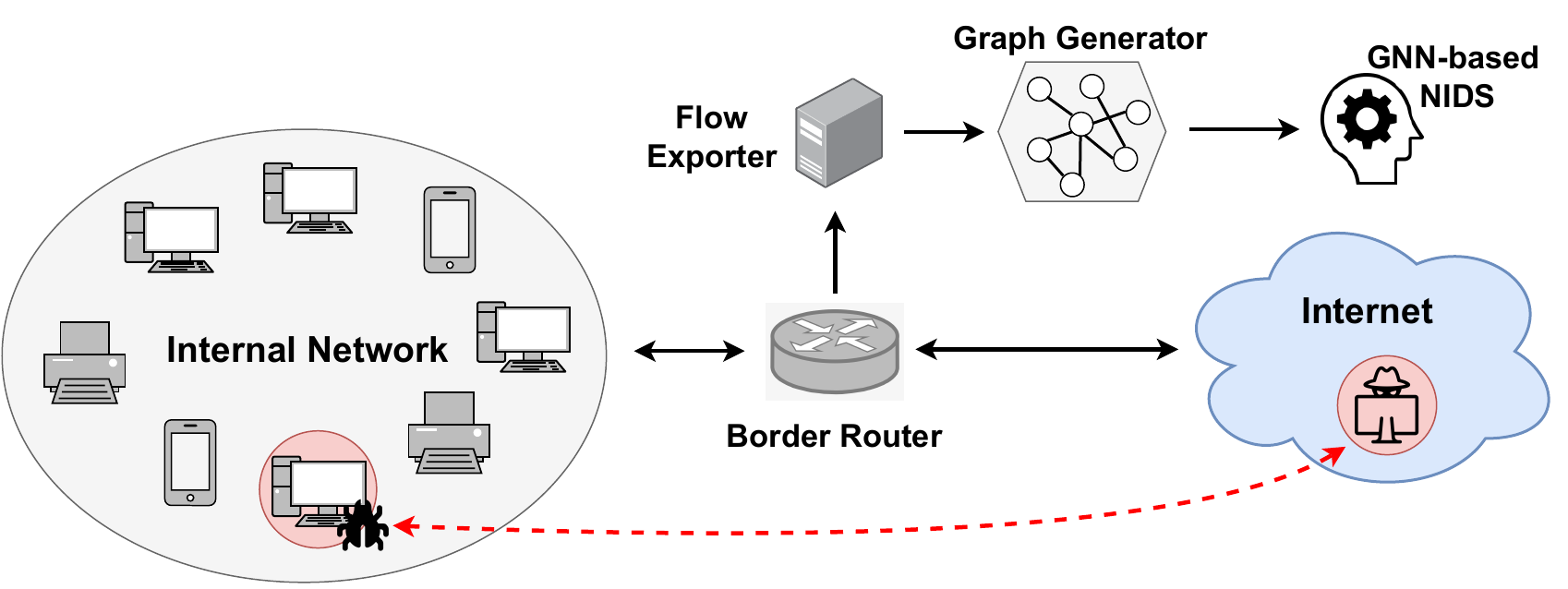}
    \caption{Threat model.}
    \label{fig:threat_model}
\end{figure}

We consider a traditional network intrusion detection scenario commonly accepted by related literature as depicted in Figure~\ref{fig:threat_model}~\cite{apruzzese2022modeling}. An attacker has already compromised at least one machine in an enterprise network formed by several hosts. The internal network is connected through a central border router that allows all the routing operations for all the hosts. This last component also serves as a proxy for sending packets to a \textit{flow exporter} in charge of extracting the netflows. Once the netflows have been obtained, another logical component called \textit{graph generator} computes the graph representation, which is, in turn, submitted to the GNN-based NIDS for classification.

The goal of the attacker is to cause evasions at test time for its malicious communications without losing the effectiveness of the original attack. For this, two strategies are possible: evasion by perturbing the feature values of their malicious netflows (feature attacks - Section~\ref{sec:feature_attacks}) or by altering the test graph itself (structural attacks - Section~\ref{sec:structural_attacks}).

We consider a realistic gray box setting for our adversarial attacks. The attacker assumes that the network is monitored by a GNN-based NIDS, but they do not know the internal parameters and settings. The complete set of features is unknown to the attacker, even if they can realistically suppose that the most important time and data-related features are employed by the classifiers, as in related work~\cite{wu2019evading, apruzzese2018evading, chernikova2022fence}. For the sake of structural attacks proposed in this paper, we assume that the attacker at least knows some of the endpoints (i.e., nodes) belonging to the monitored network. As we suppose that the attacker has already compromised some hosts of the network, we consider realistic that they also gained this kind of information. However, we also assume that the attacker does not have access to either the flow exporter, the graph generator, or the GNN-based NIDS. 

\subsection{Feature Attacks}
\label{sec:feature_attacks}
The classical strategies for generating adversarial attacks against ML-based NIDS consider perturbations to the features employed by the ML classifiers. Netflows features are still involved in the decision process of GNN-based NIDS. Hence, this type of approach is applicable also against these models. Feature attacks against ML-based NIDS have been the object of a vast line of research~\cite{apruzzese2020deep, papadopoulos2021launching}, and some attempts have also been conducted against GNN-based NIDS~\cite{pujol2022unveiling, zhou2021hierarchical}. As the main contribution of this paper is the novel formalization of structural attacks, and because feature attacks for GNN-based NIDS follow the same rules as those against traditional ML-based NIDS, a detailed discussion of their strategies is beyond the scope of this paper.
Nevertheless, we refer the reader interested in more detailed studies on feature attacks to dedicated surveys (e.g.,~\cite{apruzzese2022modeling, de2019adversarial, rosenberg2021adversarial, pierazzi2020intriguing, apruzzese2022modeling}).

\subsection{Structural attacks}
\label{sec:structural_attacks}
\begin{figure*}
    \centering
    \includegraphics[width=\textwidth]{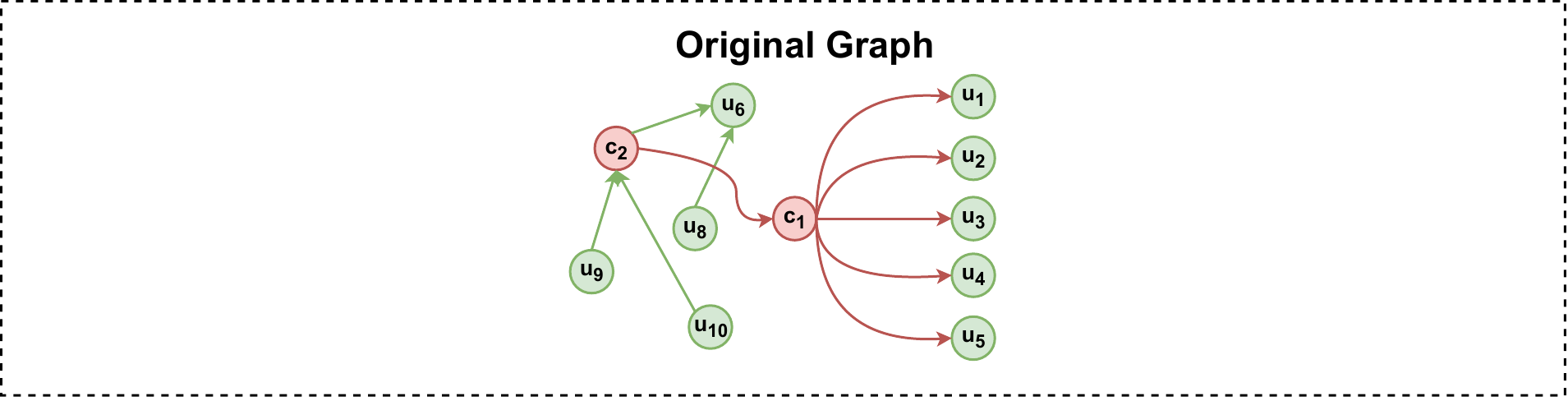}
    \caption{Flow graph example for a network scanning attack.}
    \label{fig:original_graph}
\end{figure*}

As detailed in Section~\ref{sec:background}, the embedding phase of GNN-based NIDS encapsulates netflow features and topological information about a node's neighborhood within a single vector. As a consequence, the model can detect cyberattacks not only based on netflow features, but also on the topological patterns they exhibit. A simplified example of these patterns for a network scan attack in a flow graph is depicted in Figure~\ref{fig:original_graph}. In the graph, red nodes and arrows denote compromised hosts and malicious netflows, respectively, while the green ones represent uncompromised hosts and legitimate netflows. In this case, the compromised node $c_1$ is sending scan netflows to a large set of nodes within the network, forming a significant topological pattern. The GNN-based NIDS is designed to identify such patterns, eventually flagging as malicious all netflows (edges in the flow graph) associated with them.

\begin{figure*}[ht]
  \centering
  \begin{subfigure}{.3\linewidth}
    \centering
    \includegraphics[width = \linewidth]{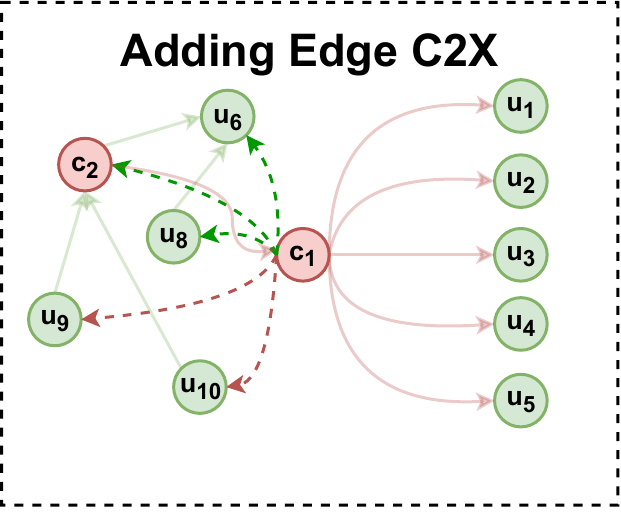}
    \caption{Example of C2x attacks.}
    \label{fig:c2x}
  \end{subfigure}%
  \hspace{1em}
  \begin{subfigure}{.3\linewidth}
    \centering
    \includegraphics[width = \linewidth]{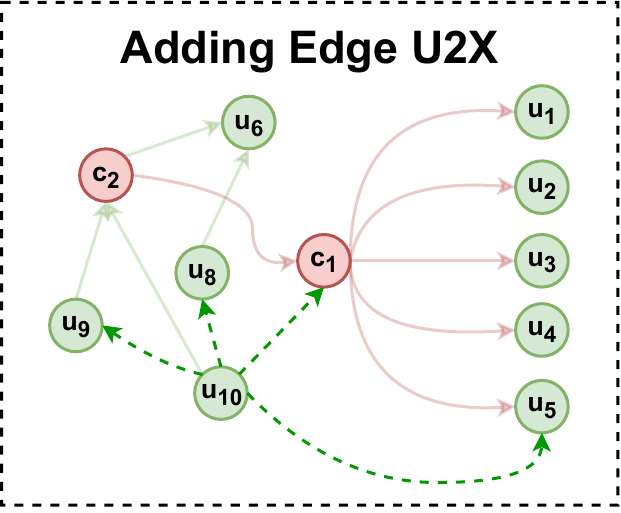}
    \caption{Example of U2x attacks.}
    \label{fig:u2x}
  \end{subfigure}%
  \hspace{2em}
  \begin{subfigure}{.3\linewidth}
    \centering
    \includegraphics[width = \linewidth]{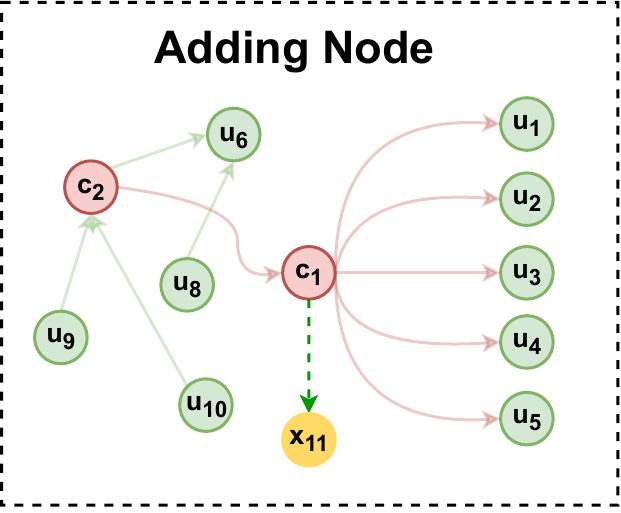}
    \caption{Example of adding node attacks.}
    \label{fig:add_node}
  \end{subfigure}
  \caption{Feasible structural adversarial attacks in practice.}
  \label{fig:aa_structural}
\end{figure*}

By perturbing the graph, the attackers can alter the neighborhoods of their cyber attacks, compromising the detection. We call this approach \textit{structural attack}. The goal is to introduce graph perturbations generating adversarial embeddings that fool the GNN-based NIDS into producing misclassifications. Notably, here, adversaries do not directly modify the malicious communications they seek to conceal; instead, they manipulate the graph structure itself (indirect attack~\cite{takahashi2019indirect, zugner2018adversarial}).    

In this paper, we propose the first formalization of \textit{structural adversarial attacks} against GNN-based NIDS.
Indeed, various structural perturbations can be applied to a graph: 
\begin{itemize}
    \item \textbf{Edge addition}: Injection of a new edge in the graph.
    \item \textbf{Edge deletion}: Removal of an edge from the graph.
    \item \textbf{Node addition}: Injection of a new node into the graph.
    \item \textbf{Node deletion}: Removal of a node from the graph.
    \item \textbf{Rewiring}: Achieved by two consecutive operations of an edge deletion and edge addition. 
\end{itemize}
We observe that some of these perturbations are feasible only if the attacker has \textit{direct access} to the graph submitted to the classifier. This is hard to achieve in realistic scenarios of network intrusion detection. In this paper, we are interested in practical solutions considering adversaries that do not have access to highly protected components, such as the graph generator or the GNN-based NIDS. 
Hence, in the next subsection, we formally present the problem space constraints limiting the attackers into operating at a traffic level, which is within their capabilities as described in our threat model.

\subsection{Structural attacks in problem space}
\label{sec:problem_space_structural}

The rationale behind our proposal is that the extraction of the netflows and their collection to generate the graph are two separate and consecutive procedures that are highly correlated~\cite{venturi2023practical}. In other words, the building of the graph submitted to the GNN-based NIDS at test time and the production of the relative predictions are operations happening \textit{after} the collection of some netflows. Hence, by operating on the network traffic, the attacker can affect the final graph submitted to the GNN-based NIDS.
Following the formalization proposed in~\cite{pierazzi2020intriguing}, we provide the first definition of the problem constraints essential for the development of structural adversarial attacks against GNN-based NIDS in realistic network intrusion detection scenarios.

\subsubsection{Available transformations} The attacker is constrained to feasible perturbations involving only addition operations on the graph (Figure~\ref{fig:aa_structural}). For instance, the attacker can introduce new edges in a flow graph by generating additional netflows (\textit{Edge addition} attacks). 
The general idea is to set up novel communications between the nodes of a network, resulting in additional packets that, once captured by the flow exporter, are transformed into netflows that will be subsequently included in the graph. 
\textit{Node addition} attacks can be considered a special case of edge addition attacks. Specifically, the attacker can introduce new nodes into the graph by exchanging packets with an endpoint that is not part of the existing graph.
On the other hand, removal transformations require a powerful attacker having access to either the flow exporter or the graph generator to manually remove specific netflows. As these components are typically among the most protected elements in the entire network, we do not consider removal attacks as feasible. We recall that our terminology refers to flow graph representations. However, it is important to remark that if flow graphs are transformed into line graphs, the edge addition attacks applied to the former are reflected in node addition attacks in line graphs thanks to the linearization procedure (Section~\ref{sec:background}).   

\subsubsection{Preserved semantics} The goal of the attacker is to cause misclassifications for the malicious communications without compromising their functionalities. Different from the case of feature attacks, here, this aspect plays a less prominent role: since the alterations to the graph do not affect the feature values of the original malicious communications, their functionalities remain preserved by construction.  

\subsubsection{Plausibility} The added entities need to appear realistic upon manual inspection. Hence, the attacker is constrained to introduce only plausible netflows that mimic authentic communications. We remark that a single packet is sufficient for modern flow exporters to generate a netflow~\cite{argus}. Hence, attackers can even send a single packet to generate a plausible netflow that will be included in the graph.

\subsubsection{Robustness to preprocessing} The attacker must account for possible non-ML mechanisms that can disrupt the attack. For instance, as we will detail in Section~\ref{sec:u2x_attack}, the addition of an edge from an uncompromised node requires the attacker to spoof the source IP address of a packet. Nevertheless, this may be thwarted by anti-spoofing techniques. Similarly, some countermeasures exist against node addition attacks. However, as we will discuss in Section~\ref{sec:add_node}, implementing such mechanisms is extremely complex in practice. 

\subsection{Proposed attacks}
\label{sec:proposed_attacks}
We propose four different structural adversarial attacks that can be performed against GNN-based NIDS in problem space.
All our attacks are variants of the adding edge attacks, and they can be modeled as a triple $(\sigma, \delta, \tau)$, in which $\sigma$ refers to the source node of the inserted edge, $\delta$ indicates the destination node (eventually a new node for adding node attacks), and $\tau$ identifies the type of traffic sent (i.e., the netflow features).

The proposed attacks are resumed in Table~\ref{tab:proposed_attacks}, and detailed in the next subsections. In the Table, for each attack, we report the conditions required for launching the attack, the type of perturbations involved, the feasibility in practice, and a brief description. We differentiate between two possible types of nodes in a flow graph. We indicate with a letter \textbf{C} a node that has already been compromised by the attacker. Conversely, with the letter \textbf{U}, we refer to uncompromised nodes. Hence, we separate two possible edge addition attacks according to the type of source node for the inserted edge. If $\sigma$ is a compromised node, we refer to a $C2x$ attack. Conversely, we indicate with $U2x$ the case in which $\sigma$ is an uncompromised host. In both cases, the letter $x$ indicates that $\delta$ is a random node in the network.

\begin{table}[htbp]
    \centering
    \caption{The proposed problem-space attacks.}
    \label{tab:proposed_attacks}
    \resizebox{\textwidth}{!}{
    \begin{tabular}{|c|c|cc|c|c|}
        \hline
        \rowcolor[HTML]{EFEFEF}
        &  & \multicolumn{2}{c|}{\textbf{Perturbation type}} &  &  \\ \cline{3-4}
        \rowcolor[HTML]{EFEFEF}
        \multirow{-2}{*}{\textbf{Attack}} & \multirow{-2}{*}{\textbf{Conditions}} & \multicolumn{1}{c|}{\textit{Feature}} & \textit{Structure} & \multirow{-2}{*}{\textbf{Feasibility}} & \multirow{-2}{*}{\textbf{Description}} \\ \hline \hline
        
        \CtoXB & \begin{tabular}[c]{@{}c@{}}$\sigma$ = \textbf{C}\\ $\delta$ = $x$\\ $\tau$ = $\mathbf{\mathcal{B}}$\end{tabular} & \multicolumn{1}{c|}{\xmark} & \checkmark & \fullcirc & \begin{tabular}[c]{@{}c@{}}Adding edge attack\\ Strategy: Attackers set up new benign communications $\mathbf{\mathcal{B}}$\\ from a compromised node to a random destination in the graph.\\ Feasible in problem space.\end{tabular} \\ \hline
        
        \CtoXM & \begin{tabular}[c]{@{}c@{}}$\sigma$ = \textbf{C}\\ $\delta$ = $x$\\ $\tau$ = $\mathbf{\mathcal{M}}$\end{tabular} & \multicolumn{1}{c|}{\xmark} & \checkmark & \fullcirc & \begin{tabular}[c]{@{}c@{}}Adding edge attack\\ Strategy: Attackers set up new malicious communications $\mathbf{\mathcal{M}}$\\ from a compromised node to a random destination in the graph.\\ Feasible in problem space.\end{tabular} \\ \hline
        
        $U2x$ & \begin{tabular}[c]{@{}c@{}}$\sigma$=\textbf{U}\\ $\delta$=$x$\\ $\tau$ = $\mathbf{\mathcal{B}}$\end{tabular} & \multicolumn{1}{c|}{\xmark} & \checkmark & \halfcirc & \begin{tabular}[c]{@{}c@{}}Adding edge attack\\ Strategy: Attackers set up new communications\\ with a spoofed source IP address corresponding to that of\\ an uncompromised node.\\ Can be hard in problem space if anti-spoofing mechanisms are present.\end{tabular} \\ \hline
        
        \textit{Add node} & \begin{tabular}[c]{@{}c@{}}$\sigma$ = \textbf{C}/\textbf{U}\\ $\delta$ = \textit{new node}\\ $\tau$ = $\mathbf{\mathcal{B}}$/$\mathbf{\mathcal{M}}$\end{tabular} & \multicolumn{1}{c|}{\xmark} & \checkmark & \fullcirc & \begin{tabular}[c]{@{}c@{}}Adding node attack\\ Strategy: Attackers set up new communications\\ with a new node that does not appear in the original graph.\\ Feasible in problem space.\end{tabular} \\ \hline
    \end{tabular}
    }
\end{table}

\subsubsection{C2x attack}
\label{sec:c2x_attack}
$C2x$ attacks represent a straightforward strategy for an attacker to produce perturbations to the graph submitted to a GNN-based NIDS by adding an additional edge from a compromised node. This is possible because the adversary can control everything about compromised nodes and can send additional traffic to generate full communications with any destination host. 

As the embeddings used by the GNN-based NIDS also encapsulate the features of the netflows in a given neighborhood, we separate two $C2x$ variants according to the type of traffic sent:
\begin{itemize}
    \item \textbf{C2x with benign traffic} (\CtoXB): the attacker sets up an additional legitimate communication with the destination. The resulting packets are collected by the flow exporter to generate a benign netflow that will be added to the graph as a new edge. The purpose of the attacker is to conceal the malicious communications coming from the compromised nodes, by injecting benign edges into their neighborhood. In this way, the embeddings produced by cyberattacks are more similar to those of legitimate communications and will be harder to detect. 
    \item \textbf{C2x with malicious traffic} (\CtoXM): the attacker sends malicious traffic to a random, possibly uncompromised, host with the goal of injecting malicious netflows into their neighborhood. Following this approach may lead to two possible outcomes. First, given that most of the nodes usually produce just legitimate traffic, the additional malicious netflows inserted into their neighborhood can be concealed. Second, their injections may cause misclassification of benign netflows, as the resulting adversarial embeddings for their neighborhoods would be similar to those of malicious communications. This is significant because a misclassification of benign netflows produces high rates of false alarms, causing a de facto Denial of Service.
\end{itemize}

An example of these two attacks can be observed in Figure~\ref{fig:c2x}. Here, the attacker is trying to deceive the GNN-based NIDS to misclassify the network scan of the compromised $c_1$ node by following the two C2x variants previously described. Legitimate communications are introduced from $c_1$ to nodes $c_2$, $u_6$ and $u_8$. This will add into the neighborhood of $c_1$ new benign netflows and uncompromised nodes that may fool the GNN-based NIDS into not recognizing the scan attack. Similarly, the attacker sends malicious communications from node $c_1$ to the uncompromised nodes $u_9$ and $u_{10}$, which are only involved in benign communications. This would further complicate the detection by the GNN-based NIDS, possibly causing a misclassification of the edges in the neighborhood of node $u_9$ and $u_10$. 

\subsubsection{U2X attack}
\label{sec:u2x_attack}

$U2x$ attacks add new edges from uncompromised nodes. This attack is more complex to achieve in practice with respect to $C2x$, as attackers do not have control of uncompromised nodes. However, source IP spoofing can make even this type of attack possible. Considering TCP connections, the workflow for this attack is as follows. From a controlled host, the attacker sends a SYN packet to a random destination in the network, crafted in order to contain a spoofed source IP address correspondent to that of an uncompromised host. The destination receives the SYN packet, and sends a SYN/ACK to the node with the spoofed IP address, which will most likely be ignored. Nevertheless, even these two packets will be enough for the flow exporter to produce the related netflow with the spoofed IP address as the source. This netflow will consequently be included in the graph generation process as a new edge. 

We highlight that if the destination of the new netflow is a compromised host, the attackers can produce sophisticated versions of the $U2x$ attack. For example, they can set the correct ACK numbers into the packets, or they can even instruct the host not to drop the spoofed packets. This allows the simulation of full communications, resulting in netflows resembling those relative to normal ones.

An example of this attack can be observed in Figure~\ref{fig:u2x}. Here, the attacker has crafted malicious packets spoofing the source IP address of the uncompromised node $u_{10}$ to generate edges to $u_5$, $u_8$, $u_9$ and $c_1$. In this case, this last edge is particularly significant as it will inject a benign netflow into the neighborhood of $c_1$, promoting a misclassification of the netflows involved in the scan attack.  

Although $U2x$ attacks can be used to alter the neighborhood of any node, they are possible only through IP spoofing. Hence, it is important to remark that anti-spoofing mechanisms can completely disrupt the efficacy of this attack. 
As these techniques are extremely common in modern networks, we will not consider this type of attack in our experimental campaign.

\subsubsection{Add node attack}
\label{sec:add_node}
Add node attacks represent special cases of the previous adding edge attacks. The workflow remains the same as in \CtoXB, \CtoXM and $U2x$. However, here, the destination of the new communication is set to a new node that does not belong to the graph. This is not complex to achieve. For example, if the attacker has some knowledge of the nodes of the graph, they can set the destination of the new packets to a host from outside the monitored network or previously not included. Moreover, as some GNN-based NIDS consider a node for each pair of IP addresses and port numbers present in the netflows~\cite{lo2022graphsage,  venturi2023arganids}, another possible strategy consists of setting up a new communication with an unused port of a selected host. Moreover, assuming that the destination node is a listening server, the attackers can also set up full communications with it.

An example of this attack is reported in Figure~\ref{fig:add_node}, in which the attacker starts a legitimate communication from the compromised node $c_1$ to $x_{11}$ that was previously not present in the graph. In this way, as in \CtoXB attacks, the purpose is to alter the neighborhood of netflows relative to the scan attack, injecting new benign netflows.

We remark that it is complex to define filtering rules for preventing add node attacks. The graph generator could exclude netflows from external nodes, but this would severely hinder the identification of external intrusions. Similarly, it is hard to know in advance the set of ports used in the monitored network to exclude all the netflows with port numbers that do not belong to it.   
\section{Testbed}
\label{sec:testbed}

As an additional contribution, we present an application of the proposed structural adversarial attacks against state-of-the-art GNN-based NIDS. In the following, we describe the classifiers and the datasets, and we detail the implementation of the adversarial attacks we consider. We release the source code for reproducing our experiments at \url{https://github.com/andreaventuri01/Structural_attacks_GNN_NIDS}.

\subsection{Classifiers}
\label{sec:classifiers}

For a comprehensive validation of the proposed adversarial attacks, we consider two GNN-based NIDS that utilize the main graph representations commonly employed in the literature for network traffic, which are flow graph and line graph (refer to Section~\ref{sec:gnn_nids}). Considering these two representations enables us to evaluate which graph representation is more vulnerable to modifications in the graph structure. We focus exclusively on \textit{inductive} models, as they possess the capability to predict classes for nodes in unseen test graphs. This is crucial for the development of realistic NIDS suitable for practical deployment. Conversely, we exclude transductive GNNs, as they necessitate the complete graph structure before starting training, which is an impractical requirement in production environments. We remark that this is a limitation of transductive approaches, not of the proposed structural attacks. Finally, we consider binary classification problems by training the GNN to detect specific attack variants in the datasets. This method aligns with recommendations in the literature for achieving high-performance classifiers tailored to each attack~\cite{apruzzese2018effectiveness, calzavara2019adversarial, wu2019evading}.   

The first GNN-based NIDS under consideration is E-GraphSAGE~\cite{lo2022graphsage}, which is based on the well-known GraphSAGE algorithm~\cite{hamilton2017inductive}. This model accepts a \textit{flow graph} as input. 
We use the public implementation of E-GraphSAGE made available by the same authors in~\cite{waimorris2022egraphsage}. There, the authors consider each unique pair of IP addresses and port number couples appearing in the netflows of the dataset as an endpoint. On the other hand, edges within the graph represent netflows. Each edge connects nodes that pertain to a netflow and stores the features of the communication. In this context, the network intrusion detection task is transformed into an edge classification problem. 

We also consider a NIDS based on the original GraphSAGE algorithm~\cite{hamilton2017inductive}, which operates over \textit{line graphs}. We refer to this model as LineGraphSAGE. As discussed in Section~\ref{sec:gnn_nids}, line graphs are derived from flow graphs through a linearization procedure. 
Utilizing a line graph for network intrusion detection translates the task from edge classification to a node classification problem. We remark that GNN-based NIDS leveraging LineGraphSAGE has not been proposed in literature per se, but it has been employed as a benchmark in assessing the performance of novel proposals in related papers (e.g., ~\cite{hu2023towards, khoury2023jbeil}). We implement LineGraphSAGE using the Pytorch Geometric~\cite{Fey/Lenssen/2019} and the Deep Graph~\cite{wang2019dgl} libraries.

\subsection{Datasets}
\label{sec:datasets}
In our experimental campaign, we use two publicly available datasets for network intrusion detection: \textbf{CTU-13}~\cite{garcia2014empirical} and \textbf{ToN-IoT}~\cite{alsaedi2020ton_iot}. These datasets consist of real traffic samples and have been previously employed in related literature for GNN-based NIDS~\cite{bilot2023graph, venturi2023arganids, lo2022graphsage}. CTU-13 comprises traffic captures in which several botnet variants are executed in a medium-sized network. Conversely, the ToN-IoT dataset contains traffic traces from various IoT devices encompassing multiple types of attacks beyond botnets. Overall, the malicious activities considered in the two datasets expose different structural patterns and represent good candidates for assessing GNN-based NIDS.

We apply standard preprocessing operations~\cite{venturi2021drelab} to make data suitable for the GNN-based NIDS. 
Considering the binary detection problem that we tackle in this paper, we create one different data collection for each attack in the datasets. Hence, we divide the data collections into training and test sets by considering 80:20 train-test splits. Moreover, each set is formed by merging together malicious samples from a specific attack and benign netflows in a 1:10 ratio. This allows us to maintain a data unbalancing that is more similar to real-world scenarios~\cite{apruzzese2018evading}.

In Table~\ref{tab:ctu_distrib} and~\ref{tab:ton_distrib}, we report more specific information for the datasets considered in our evaluation campaign. 

\begin{table}[htbp]
\centering
\caption{Train and test flow graph structure for each attack in the CTU dataset.}
\label{tab:ctu_distrib}
\resizebox{0.7\linewidth}{!}{
    \begin{tabular}{|c||ccc||ccc|}
        \hline
        \rowcolor[HTML]{EFEFEF} 
        \cellcolor[HTML]{EFEFEF} & \multicolumn{3}{c||}{\cellcolor[HTML]{EFEFEF}\textbf{Train}} & \multicolumn{3}{c|}{\cellcolor[HTML]{EFEFEF}\textbf{Test}} \\ \cline{2-7} 
        \rowcolor[HTML]{EFEFEF} 
        \multirow{-2}{*}{\cellcolor[HTML]{EFEFEF}\textbf{Attack}} & \multicolumn{1}{c|}{\cellcolor[HTML]{EFEFEF}\textit{\#Mal}} & \multicolumn{1}{c|}{\cellcolor[HTML]{EFEFEF}\textit{\begin{tabular}[c]{@{}c@{}}\#Nodes\\ (\#C)\end{tabular}}} & \textit{\#Edges} & \multicolumn{1}{c|}{\cellcolor[HTML]{EFEFEF}\textit{\#Mal}} & \multicolumn{1}{c|}{\cellcolor[HTML]{EFEFEF}\textit{\begin{tabular}[c]{@{}c@{}}\#Nodes\\ (\#C)\end{tabular}}} & \textit{\#Edges} \\ \hline \hline
        
        \textit{Neris} & \multicolumn{1}{c|}{$60\,854$} & \multicolumn{1}{c|}{\begin{tabular}[c]{@{}c@{}}$545\,286$\\ ($27\,977$)\end{tabular}} & $14\,200\,942$ & \multicolumn{1}{c|}{$20\,284$} & \multicolumn{1}{c|}{\begin{tabular}[c]{@{}c@{}}$216\,225$\\ ($13\,755$)\end{tabular}} & $473\,646$ \\ \hline
        \textit{Rbot} & \multicolumn{1}{c|}{$20\,632$} & \multicolumn{1}{c|}{\begin{tabular}[c]{@{}c@{}}$346\,256$\\ ($4\,002$)\end{tabular}} & $797\,008$ & \multicolumn{1}{c|}{$6\,877$} & \multicolumn{1}{c|}{\begin{tabular}[c]{@{}c@{}}$134\,297$\\ ($3\,404$)\end{tabular}} & $265\,670$ \\ \hline
        \textit{Virut} & \multicolumn{1}{c|}{$5\,392$} & \multicolumn{1}{c|}{\begin{tabular}[c]{@{}c@{}}$113\,891$\\ ($4\,841$)\end{tabular}} & $226\,462$ & \multicolumn{1}{c|}{$1\,797$} & \multicolumn{1}{c|}{\begin{tabular}[c]{@{}c@{}}$42\,788$\\ ($1\,739$)\end{tabular}} & $75\,486$ \\ \hline
        \textit{Menti} & \multicolumn{1}{c|}{$611$} & \multicolumn{1}{c|}{\begin{tabular}[c]{@{}c@{}}$16\,269$\\ ($564$)\end{tabular}} & $25\,672$ & \multicolumn{1}{c|}{$204$} & \multicolumn{1}{c|}{\begin{tabular}[c]{@{}c@{}}$5\,863$\\ ($201$)\end{tabular}} & $8\,558$ \\ \hline
        \textit{Murlo} & \multicolumn{1}{c|}{$800$} & \multicolumn{1}{c|}{\begin{tabular}[c]{@{}c@{}}$20\,403$\\ ($782$)\end{tabular}} & $33\,596$ & \multicolumn{1}{c|}{$266$} & \multicolumn{1}{c|}{\begin{tabular}[c]{@{}c@{}}$7\,453$\\ ($262$)\end{tabular}} & $11\,198$ \\ \hline
    \end{tabular}
}
\end{table}

\begin{table}[htbp]
\centering
\caption{Train and test flow graphs structure for each attack in the ToN-IoT dataset.}
\label{tab:ton_distrib}
\resizebox{0.7\linewidth}{!}{
    \begin{tabular}{|c||ccc||ccc|}
        \hline
        \rowcolor[HTML]{EFEFEF} 
        \cellcolor[HTML]{EFEFEF} & \multicolumn{3}{c||}{\cellcolor[HTML]{EFEFEF}\textbf{Train}} & \multicolumn{3}{c|}{\cellcolor[HTML]{EFEFEF}\textbf{Test}} \\ \cline{2-7} 
        \rowcolor[HTML]{EFEFEF} 
        \multirow{-2}{*}{\cellcolor[HTML]{EFEFEF}\textbf{Attack}} & \multicolumn{1}{c|}{\cellcolor[HTML]{EFEFEF}\textit{\#Mal}} & \multicolumn{1}{c|}{\cellcolor[HTML]{EFEFEF}\textit{\begin{tabular}[c]{@{}c@{}}\#Nodes\\ (\#C)\end{tabular}}} & \textit{\#Edges} & \multicolumn{1}{c|}{\cellcolor[HTML]{EFEFEF}\textit{\#Mal}} & \multicolumn{1}{c|}{\cellcolor[HTML]{EFEFEF}\textit{\begin{tabular}[c]{@{}c@{}}\#Nodes\\ (\#C)\end{tabular}}} & \textit{\#Edges} \\ \hline \hline
        \textit{Bkdr} & \multicolumn{1}{c|}{$14\,992$} & \multicolumn{1}{c|}{\begin{tabular}[c]{@{}c@{}}$47\,851$\\ ($788$)\end{tabular}} & $317\,192$ & \multicolumn{1}{c|}{$4\,998$} & \multicolumn{1}{c|}{\begin{tabular}[c]{@{}c@{}}$20\,857$\\ ($701$)\end{tabular}} & $105\,677$ \\ \hline
        \textit{DDoS} & \multicolumn{1}{c|}{$11\,048$} & \multicolumn{1}{c|}{\begin{tabular}[c]{@{}c@{}}$47\,634$\\ ($9\,856$)\end{tabular}} & $233\,936$ & \multicolumn{1}{c|}{$3\,682$} & \multicolumn{1}{c|}{\begin{tabular}[c]{@{}c@{}}$19\,401$\\ ($3\,546$)\end{tabular}} & $77\,834$ \\ \hline
        \textit{DoS} & \multicolumn{1}{c|}{$13\,890$} & \multicolumn{1}{c|}{\begin{tabular}[c]{@{}c@{}}$45\,631$\\ ($271$)\end{tabular}} & $293\,772$ & \multicolumn{1}{c|}{$4\,630$} & \multicolumn{1}{c|}{\begin{tabular}[c]{@{}c@{}}$19\,807$\\ ($202$)\end{tabular}} & $97\,899$ \\ \hline
        \textit{Inj} & \multicolumn{1}{c|}{$14\,452$} & \multicolumn{1}{c|}{\begin{tabular}[c]{@{}c@{}}$57\,760$\\ ($11\,767$)\end{tabular}} & $305\,895$ & \multicolumn{1}{c|}{$4\,818$} & \multicolumn{1}{c|}{\begin{tabular}[c]{@{}c@{}}$24\,400$\\ ($4\,493$)\end{tabular}} & $101\,883$ \\ \hline
        \textit{Psw} & \multicolumn{1}{c|}{$14\,865$} & \multicolumn{1}{c|}{\begin{tabular}[c]{@{}c@{}}$59\,346$\\ ($12\,118$)\end{tabular}} & $314\,371$ & \multicolumn{1}{c|}{$4\,955$} & \multicolumn{1}{c|}{\begin{tabular}[c]{@{}c@{}}$24\,638$\\ ($4\,598$)\end{tabular}} & $104\,800$ \\ \hline
        \textit{Rans} & \multicolumn{1}{c|}{$14\,830$} & \multicolumn{1}{c|}{\begin{tabular}[c]{@{}c@{}}$48\,456$\\ ($1\,393$)\end{tabular}} & $313\,653$ & \multicolumn{1}{c|}{$4\,944$} & \multicolumn{1}{c|}{\begin{tabular}[c]{@{}c@{}}$21\,232$\\ ($1\,285$)\end{tabular}} & $104\,667$ \\ \hline
        \textit{Scan} & \multicolumn{1}{c|}{$14\,568$} & \multicolumn{1}{c|}{\begin{tabular}[c]{@{}c@{}}$51\,846$\\ ($1\,274$)\end{tabular}} & $308\,190$ & \multicolumn{1}{c|}{$4\,856$} & \multicolumn{1}{c|}{\begin{tabular}[c]{@{}c@{}}$22\,943$\\ ($599$)\end{tabular}} & $102\,706$ \\ \hline
        \textit{XSS} & \multicolumn{1}{c|}{$13\,950$} & \multicolumn{1}{c|}{\begin{tabular}[c]{@{}c@{}}$52\,739$\\ ($7\,795$)\end{tabular}} & $295\,078$ & \multicolumn{1}{c|}{$4\,650$} & \multicolumn{1}{c|}{\begin{tabular}[c]{@{}c@{}}$22\,695$\\ ($3\,722$)\end{tabular}} & $98\,462$ \\ \hline
    \end{tabular}
}
\end{table}
Along with the number of malicious samples in both train and test sets, we also provide the total number of nodes and edges (along with the number of compromised nodes) for the flow graphs generated with the netflows of each set. 
We remark that our flow graph representation has a node for each pair of IP addresses and port numbers. Hence, we consider a node being \textbf{compromised} by the attacker if its IP address appears as a source IP address in any malicious netflow of the dataset. 

\subsection{Adversarial attacks implementation}
\label{sec:implementation_aa}

We consider only attacks that are feasible in practice, taking into account the constraints discussed above.
In particular, we propose four different attacks, whose implementation is detailed below. 

\textbf{Feature-based attack}: this category of attacks involves perturbations to the feature of the netflows submitted to the classifier at test time. Our purpose when considering feature attacks is to verify whether GNN-based NIDS can be evaded by adversarial attacks effective against traditional ML-based NIDS without perturbing the original graph structures. We choose an attack strategy that has been shown to evade the detection provided by traditional ML-based NIDS in~\cite{apruzzese2018evading}. The idea of the attack is to manually modify combinations of up to four features in the malicious samples that are altered by fixed amounts. The selected features are \textit{duration}, \textit{inbytes}, \textit{outbytes}, and \textit{totpackets}, which attackers can easily modify. For example, they can increment the \textit{duration} of a netflow by inserting delays in their communications; similarly, they can change the \textit{number of bytes} by appending junk data; or they can add junk packets. The four features are combined into $15$ different combination groups. Each group involves one or more of the selected features, which are incremented by fixed amounts in different steps. The original paper proposes $9$ steps that span from an increment of $+1$ to $+1024$. Hence, after having selected a group and a step, the attacker modifies the features of the malicious samples in the test set belonging to the chosen group and perturbs them according to the step. The resulting modified samples are submitted to the classifier. In total, we have $135$ different adversarial samples from the test sets for each attack in the datasets. 
    
\CtoXB~\textbf{attack}: this is the first structural attack we consider in our evaluation campaign, and it is among the novel adding edge attacks introduced for the first time in this paper. Here, we simulate attackers setting up additional \textit{benign} communications from their compromised nodes to random destinations in the network. This results in novel netflows included in the test set, which will introduce new edges in the consequent flow graph. This is possible because the test graph generation happens only after the netflows in the test set have been produced and collected (Sections~\ref{sec:gnn_nids} and~\ref{sec:problem_space_structural}). To implement a feasible version of this attack, we proceed as follows. From each compromised node in the test set, we send $\beta$ new benign flows to random nodes. To do this, we randomly sample the additional benign network flows from the pool of benign flows available in the test set of each attack. This strategy allows us to simulate realistic benign communications as they appear in the original dataset. Subsequently, we substitute the source IP addresses and ports of the selected benign flows with those of the compromised nodes. This ensures that in the graph, the new edges originate from the compromised nodes. We obtain $\beta * |\bm{C}|$ new benign edges, where $|\bm{C}|$ indicates the number of compromised nodes in the test set. We consider different fixed step values of $\beta$: $1$, $2$, $5$, $10$, $20$.
    
\CtoXM~\textbf{attack}: we simulate attackers setting up novel malicious communications from their compromised nodes to random destinations in the graphs. From each compromised node, we send $\theta$ new malicious flows to random nodes. To do this, we randomly sample $\theta$ additional malicious netflows from the ones appearing in the corresponding test set. In this way, we can simulate realistic malicious communications as provided in the original dataset. Subsequently, we extract destination addresses and ports from other $\theta$ randomly extracted benign flows and substitute them into the corresponding fields of the new set of malicious flows. We do not modify the sources of the new malicious flows, as they already belong to compromised nodes (Section~\ref{sec:datasets}). In this way, we ensure that in the resulting graph, the malicious flows originate from a compromised host and are sent to a random endpoint. At the end, we obtain $\theta * |\bm{C}|$ new benign edges, where $|\bm{C}|$ indicates the number of compromised nodes in the test set. We consider different fixed step values of $\theta$: $1$, $2$, $5$, $10$, $20$.
    
\textbf{Add Node attack}: this is the last structural attack we consider in our evaluation campaign. As discussed previously, adding node attacks can be seen as a special case for adding edge attacks. Our implementation relies on the \CtoXB~attack previously introduced. Here we have two parameters: $\eta$ and $\gamma$. $\eta$ refers to the number of new nodes generated, while $\gamma$ indicates the number of new benign flows sent from compromised nodes to each of the $\eta$ new nodes. To produce realistic netflows, we proceed as follows. First, we generate a set of $\eta$ new nodes. This is simple to achieve, as it is sufficient to select combinations of IP addresses and port numbers that do not appear in the dataset. We assume that the new nodes are listening servers, allowing us to set up full communications with them. Hence, we send $\gamma$ benign flows from compromised nodes to each of the $\eta$ new nodes. This last procedure is similar to the one followed for the \CtoXB~attack. For each new node: (i) we sample $\gamma$ benign flows; (ii) we substitute their source IP addresses and ports with those of compromised nodes; (iii) we substitute their destination IP addresses and ports with those of the new nodes. At the end, we obtain $\eta * \gamma$ new benign edges. In our experimental campaign, we set $\eta$ to $1$, $5$, $10$, $100$ and $1000$, while $\gamma$ is set to $1$, $5$ and $20$.

We remark that we implement all the structural perturbations of our attacks on the flow graph representations generated from the netflows in each test set. Hence, we obtain the perturbed line graph representations by applying the linearization procedure described in Section~\ref{sec:gnn_nids} to the perturbed flow graph. The results will be submitted to the LineGraphSAGE model. 
\section{Experimental Results}
\label{sec:results}
The objective of the experimental campaign is to demonstrate the effectiveness of the proposed structural adversarial attacks against the NIDS based on E-GraphSAGE and LineGraphSAGE. For evaluating the performance in the baseline scenario, which does not involve any adversarial perturbation, we adopt the metrics largely employed in machine learning: \textit{F1-score}, \textit{Recall} (or \textit{Detection Rate - DR}), and \textit{Precision}. Conversely, for adversarial attacks, we consider the DR for the malicious samples.

\subsection{Baseline results}
\label{sec:baseline}

Table~\ref{tab:baseline} presents the result of the baseline GNN-based NIDS in standard evaluation contexts in which no perturbations are involved. We note that the results of the classifiers are in line with those at the state-of-the-art~\cite{gamage2020deep, lo2022graphsage, pujol2022unveiling}. From Table~\ref{tab:baseline}, we observe that LineGraphSAGE slightly outperforms E-GraphSAGE in both datasets with average F1-scores of $0.951$ and $0.986$ against $0.942$ and $0.978$, respectively for the CTU-13 and the ToN-IoT datasets. These results denote the superiority of the GNN-based NIDS for node classification over those for edge classification, even if the deployment of the LineGraphSAGE model is more computationally expensive, given the additional linearization procedure required. Overall, the high Recall values of these models indicate their ability to detect most malicious samples. Hence, they represent a valid benchmark for the validation of our attacks.

\begin{table}[htbp]
    \centering
    \caption{Performance of the baseline E-GraphSAGE and LineGraphSAGE in non-adversarial contexts.}
    \label{tab:baseline}
    \resizebox{\linewidth}{!}{
    \begin{tabular}{|c|c|c|c|c||c|c|c|}
    \hline
    \rowcolor[HTML]{EFEFEF}
     \multicolumn{2}{|c|}{\textbf{Detector}}  & \multicolumn{3}{c||}{\textbf{E-GraphSAGE}} & \multicolumn{3}{c|}{\textbf{LineGraphSAGE}} \\ \hline
    \rowcolor[HTML]{EFEFEF}
    \textbf{Dataset} & \textbf{Attack} & \textit{\textbf{F1-score}} & \textit{\textbf{Recall}} & \textit{\textbf{Precision}} & \textit{\textbf{F1-score}} & \textit{\textbf{Recall}} & \textit{\textbf{Precision}} \\ \hline \hline
     & \textit{Neris} & $0.859$ & $0.846$ & $0.871$ & $0.852$ & $0.787$ & $0.929$ \\ \cline{2-8} 
     & \textit{Rbot} &  $0.986$ & $0.988$ & $0.984$ & $0.989$ & $0.993$ & $0.991$ \\ \cline{2-8} 
     & \textit{Virut} & $0.919$ & $0.876$ & $0.967$ & $0.991$ & $0.882$ & $0.933$ \\ \cline{2-8} 
     & \textit{Menti} & $0.985$ & $0.998$ & $0.973$ & $0.971$ & $0.950$ & $0.992$ \\ \cline{2-8} 
     & \textit{Murlo} & $0.961$ & $0.962$ & $0.959$ & $0.954$ & $0.970$ & $0.938$ \\ \cline{2-8} 
     \multirow{-6}{*}{\rotatebox[origin=c]{90}{\textbf{CTU-13}}} & 
     \cellcolor[HTML]{EFEFEF} \cellcolor[HTML]{EFEFEF} \begin{tabular}{@{}c@{}} \textit{average} \\ (\textit{std. dev.})\end{tabular} &
     \cellcolor[HTML]{EFEFEF} \begin{tabular}{@{}c@{}}  $0.942$ \\ ($0.053$)\end{tabular} & 
     \cellcolor[HTML]{EFEFEF} \begin{tabular}{@{}c@{}} $0.934$ \\ ($0.068$)\end{tabular} & 
     \cellcolor[HTML]{EFEFEF} \begin{tabular}{@{}c@{}} $0.950$ \\ ($0.045$)\end{tabular} & 
     \cellcolor[HTML]{EFEFEF} \begin{tabular}{@{}c@{}} $0.951$ \\ ($0.057$)\end{tabular} & 
     \cellcolor[HTML]{EFEFEF} \begin{tabular}{@{}c@{}} $0.916$ \\ ($0.083$)\end{tabular} & 
     \cellcolor[HTML]{EFEFEF} \begin{tabular}{@{}c@{}} $0.956$ \\ ($0.030$)\end{tabular} \\ \hline \hline 
    
     & \textit{DDoS} & $0.976$ & $0.969$ & $0.984$ & $0.990$ & $0.985$ & $0.994$ \\ \cline{2-8} 
     & \textit{DoS} &  $0.977$ & $0.967$ & $0.987$ & $0.990$ & $0.988$ & $0.993$ \\ \cline{2-8} 
     & \textit{Bkdr} & $0.994$ & $0.993$ & $0.995$ & $0.997$ & $0.997$ & $0.998$ \\ \cline{2-8} 
     & \textit{Scan} & $0.935$ & $0.910$ & $0.963$ & $0.938$ & $0.892$ & $0.990$ \\ \cline{2-8} 
     & \textit{Rans} & $0.990$ & $0.996$ & $0.983$ & $0.995$ & $0.993$ & $0.997$ \\ \cline{2-8} 
     & \textit{Psw} &  $0.986$ & $0.976$ & $0.995$ & $0.996$ & $0.996$ & $0.996$ \\ \cline{2-8}
     & \textit{Inj} &  $0.984$ & $0.986$ & $0.982$ & $0.993$ & $0.999$ & $0.987$ \\ \cline{2-8}
     & \textit{XSS} &  $0.982$ & $0.974$ & $0.989$ & $0.996$ & $0.997$ & $0.995$ \\ \cline{2-8}
     \multirow{-9}{*}{\rotatebox[origin=c]{90}{\textbf{ToN-IoT}}} &
     \cellcolor[HTML]{EFEFEF} \cellcolor[HTML]{EFEFEF} \begin{tabular}{@{}c@{}} \textit{average} \\ (\textit{std. dev.})\end{tabular} &
     \cellcolor[HTML]{EFEFEF} \begin{tabular}{@{}c@{}}  $0.978$ \\ ($0.018$)\end{tabular} &
     \cellcolor[HTML]{EFEFEF} \begin{tabular}{@{}c@{}} $0.971$ \\ ($0.027$)\end{tabular} & 
     \cellcolor[HTML]{EFEFEF} \begin{tabular}{@{}c@{}} $0.984$ \\ ($0.010$)\end{tabular} &
     \cellcolor[HTML]{EFEFEF} \begin{tabular}{@{}c@{}} $0.986$ \\ ($0.019$)\end{tabular} &
     \cellcolor[HTML]{EFEFEF} \begin{tabular}{@{}c@{}} $0.980$ \\ ($0.036$)\end{tabular} &
     \cellcolor[HTML]{EFEFEF} \begin{tabular}{@{}c@{}} $0.993$ \\ ($0.003$)\end{tabular} \\ \hline
    
    \end{tabular}
    }
\end{table}

\subsection{Feature-based attacks}
\label{sec:feature_results}

After having confirmed the high performance offered by GNN-based NIDS in non-adversarial contexts, we test their robustness against feature-based attacks. In this case, we launch feature evasion attacks that have been shown successful against traditional ML-based NIDS in~\cite{apruzzese2018evading}. For this reason, we compare the performance of the considered GNN-based NIDS against a traditional ML-based NIDS based on the Random Forests (RF) algorithm, which is indicated as the most effective traditional ML classifier for network intrusion detection~\cite{calzavara2019adversarial, apruzzese2018effectiveness}. In particular, we train the RF model on the same train-test division described in Tables~\ref{tab:ctu_distrib} and~\ref{tab:ton_distrib}, using the same parameters as indicated in the original work that proposed the feature attack we consider~\cite{apruzzese2018evading}.  

\begin{figure}[tbp]
    \centering
    \includegraphics[width=1.3\textwidth, center]{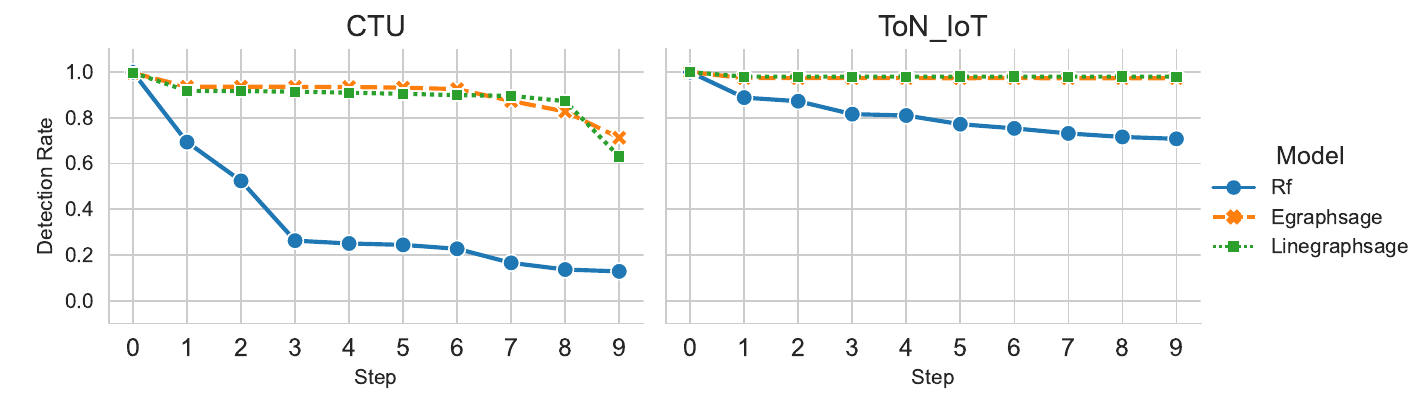}
    \caption{Feature attack against RF, E-GraphSAGE, and LineGraphSAGE.}
    \label{fig:feature_atk}
\end{figure}

The results are reported in Figure~\ref{fig:feature_atk}. For the sake of space limitations, we do not provide the detailed results for all the classifiers and for all the $135$ different adversarial attack variants. Instead, in the Figure, we provide the average results considering each binary classifier and all the feature groups of the modifications in the respective dataset. The two plots in Figure~\ref{fig:feature_atk} refer to the CTU and the ToN-IoT dataset, respectively, and show the average DR (y-axis) for the three considered NIDS, obtained by applying the perturbation according to the $9$ different incremental steps (x-axis) described in Section~\ref{sec:implementation_aa}. The Step $0$ indicates the baseline performance in non-adversarial contexts. 

From the plot referring to the CTU dataset, we can immediately note the resilience of both the GNN-based NIDS against feature attacks. This is in contrast with the extreme vulnerability shown by the RF classifier. In particular, the RF classifier experiences a significant degradation in detection performance as the number of steps increases, approaching the poor average DR score of $0.1$ in the $9$th step. Conversely, E-GraphSAGE and LineGraphSAGE maintain high DRs throughout the attack sequence, with marginal declines only when considering the largest perturbations. A similar behavior is observable also by looking at the plot for the ToN-IoT dataset. Although for this dataset the DR for the RF classifier does not drop as much as in the previous case, the superiority of the E-GraphSAGE and LineGraphSAGE classifiers is still evident. Notably, we have not experienced any performance drop for the GNN-based NIDS, even in the last steps. 

These results confirm the intuition that the structure and learning mechanisms of GNN-based models confer greater robustness against the type of feature evasion attacks that were previously successful against traditional ML classifiers like RF. Moreover, we note that there is no preferable classifier between E-GraphSAGE and LineGraphSAGE, as they offer comparable performance. 
The resilience of GNN-based NIDS to feature attacks has also been discussed in~\cite{pujol2022unveiling}. In their research, the authors focus on only two possible target features: \textit{packet size} and \textit{inter-arrival time}, with modifications being capped at an additional $200$ bytes and an extra $2$ more seconds, respectively. Our analysis, instead, covers a broader range of features and includes more significant perturbations. Our findings align with those in~\cite{pujol2022unveiling}, confirming that GNN-based NIDS exhibit greater robustness compared to traditional ML techniques. 

\subsection{\CtoXB~attack}
\label{sec:ben_from_C}

Structural attacks represent the main contribution of this paper. We now present the results of the GNN-based NIDS against \CtoXB attack. In this scenario, we inject $\beta$ new benign netflows originating from every compromised node to arbitrary targets.  

\begin{figure}[htbp]
    \centering
    \includegraphics[width=1.3\textwidth, center]{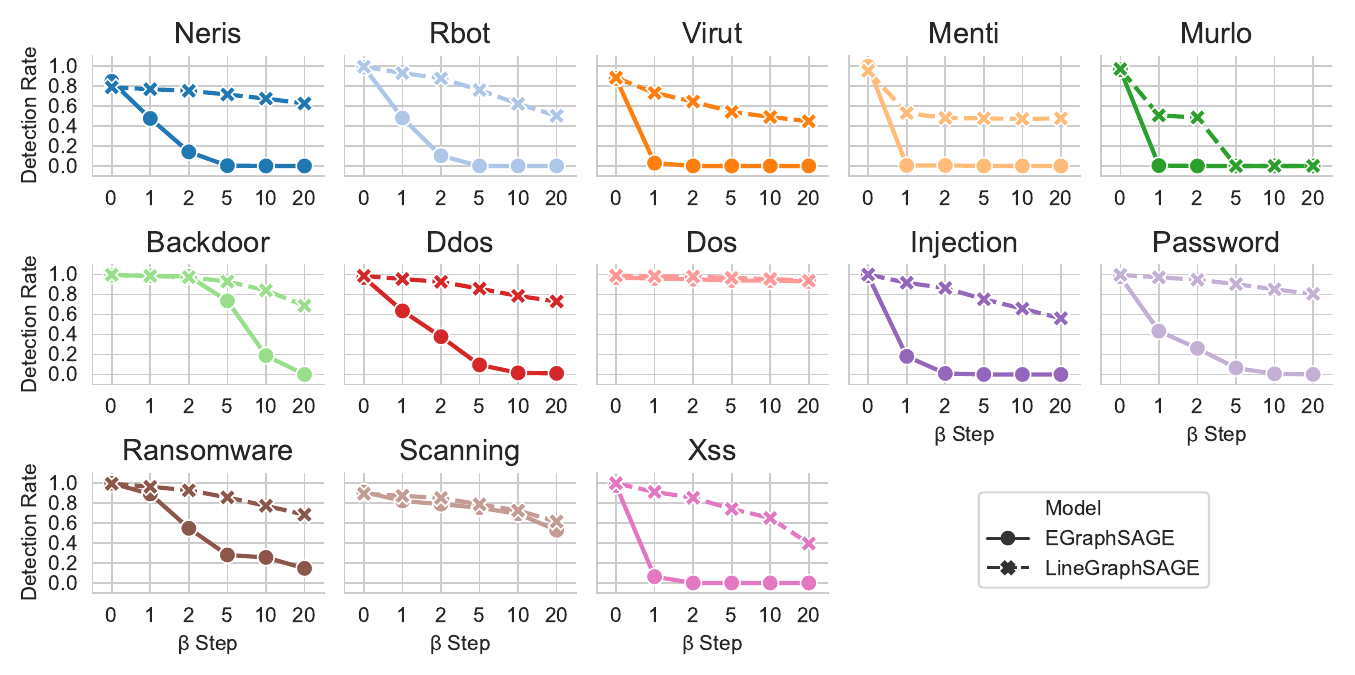}
    \caption{\CtoXB~attack against E-GraphSAGE and LineGraphSAGE.}
    \label{fig:ben_from_C}
\end{figure}

The plots in Figure~\ref{fig:ben_from_C} illustrate the DR of E-GraphSAGE and LineGraphSAGE, against the incremental insertion of the benign netflows. On the y-axis we can observe the DR, while the x-axis represents the number of $\beta$ insertions. It is important to note that when $\beta$ is set to $0$, it corresponds to the original, unperturbed graph, serving as the baseline DR for comparison. Each subplot corresponds to a different type of attack present in the datasets. 

The general downward trend in DRs as $\beta$ increases from $0$ to $20$ indicates a progressive degradation in the performance of the models. At $\beta=0$, the DRs are at their highest, reflecting their efficacy under normal, unperturbed conditions. However, as benign netflows are progressively injected into the system, we see a consistent decline in the ability of the models to accurately detect malicious activities, which is evident in the downward slopes of the lines in each plot. 
Nevertheless, the decline is not uniform across all attack types and all the models. Notably, LineGraphSAGE generally maintains a higher DR across each attack type when compared to E-GraphSAGE, suggesting its better resilience to structural perturbations. The most pronounced decreases are shown by the classifiers for the botnet variants of the CTU dataset. In particular, the E-GraphSAGE model drops to insufficient scores already when $\beta = 1$, denoting how it is sufficient for attackers to inject a single additional benign netflow from their compromised nodes to evade the detection of most malicious communications. On the other hand, the LineGraphSAGE algorithm remains more stable, although it falls below $0.5$ when $\beta > 5$ in most of the cases. Similar conclusions can be drawn by looking at the plots relative to the attacks of the ToN-IoT dataset. In most attacks, the E-GraphSAGE model shows significant losses already from the first $\beta$ steps, while LineGraphSAGE exhibits stronger - albeit insufficient - robustness. As a notable exception, we observe how the classifiers for the DoS algorithm seem not to be affected by the \CtoXB attacks, regardless of the $\beta$ value. This anomaly could be attributed to the fact that the test set of the DoS attack has the lowest number of compromised nodes (only $202$ as mentioned in Table~\ref{tab:ton_distrib}), which restricts the influence of the attacker.

\subsection{\CtoXM~attack}
\label{sec:mal_from_C}

\CtoXM represents another novel structural attack introduced in this paper. In this scenario, we send $\theta$ additional malicious netflows from each node compromised by the attackers to random targets.

\begin{figure}[htbp]
    \centering
    \includegraphics[width=1.3\textwidth, center]{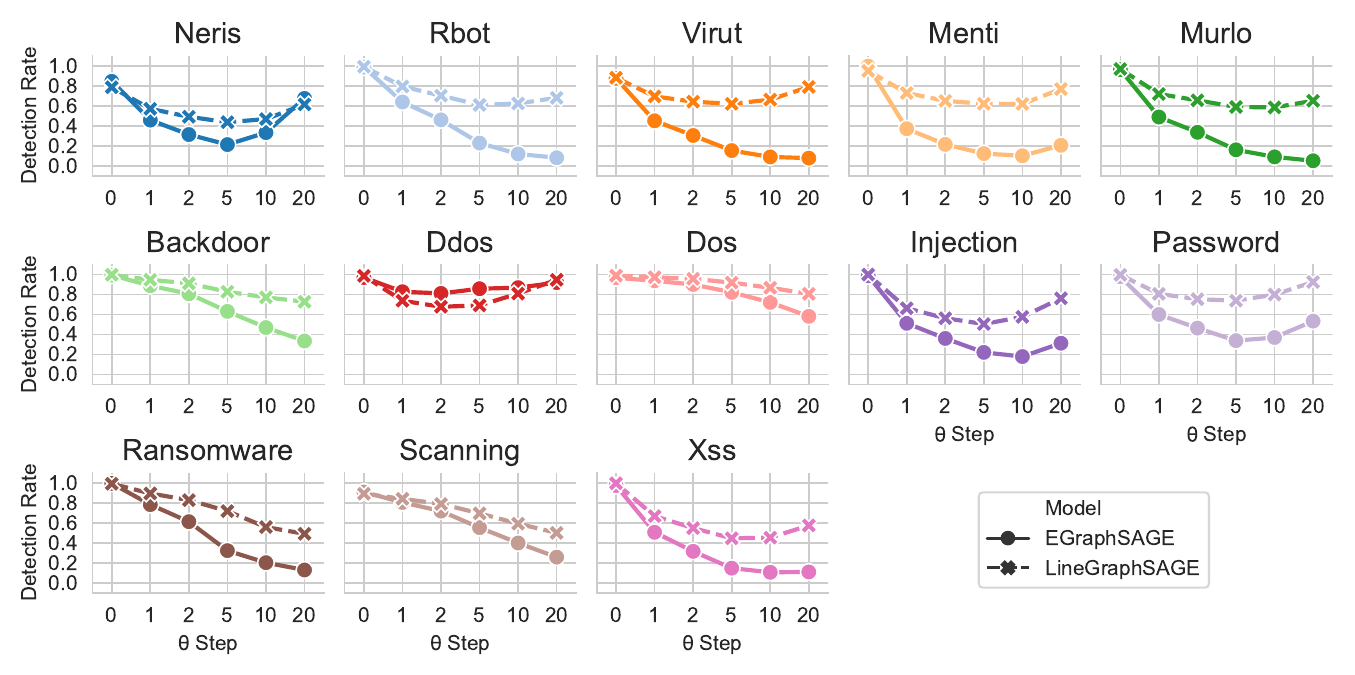}
    \caption{\CtoXM~attack against E-GraphSAGE and LineGraphSAGE.}
    \label{fig:mal_from_C}
\end{figure}

The results are reported in Figure~\ref{fig:mal_from_C}. As before, every plot includes two lines, each referring to the E-GraphSAGE and LineGraphSAGE classifiers, respectively. The y-axis shows the DR, and the x-axis displays the values we consider for $\theta$. Again, the $\theta$ value of $0$ indicates the original, unperturbed graph.

From the plots, we can observe a similar descending trend in the DRs as in the \CtoXB~scenario. However, the descending slopes for the E-GraphSAGE model are now less steep, although they still approach $0$ toward the largest $\theta$ values in most cases. LineGraphSAGE remains the most robust model, even if its resilience is still not enough to maintain sufficient DRs when increasing $\theta$. The classifiers trained on the botnet of CTU persist as the most vulnerable to structural attacks. 
As an additional note, we observe how, for the classifiers of some attack variants, there exists an unexpected trend in which the DR grows toward the largest $\theta$. This is particularly evident for the detectors trained on the Neris, Menti, Injection, Password-Cracking, and XSS attacks. We conclude that for these variants, the excessive insertion of malicious netflows leads to structural patterns that are recognized by the classifiers.    

Another interesting case is offered by the algorithms trained on the DDoS attack. Different from the severe drops in the previous case, we note that now the models remain almost stable, especially for E-GraphSAGE. A possible explanation for this behavior is due to the fact that the structural patterns for DDoS attacks involve a large number of malicious flows originating from compromised nodes. Hence, the addition of novel malicious flows from the same endpoints may not be enough to break the topological pattern exhibited by the cyberattack.

\subsection{Add Node attack}
\label{sec:results_add_node}

The last structural attack variant we consider in this paper is the Add Node attack. For this scenario, we send $\gamma$ benign netflows from compromised nodes to each of the $\eta$ new nodes not included in the original graph.

\begin{figure*}[htbp]
    \centering
    \begin{subfigure}[htbp]{\textwidth}
        \centering
        \includegraphics[width=1.3\textwidth, center]{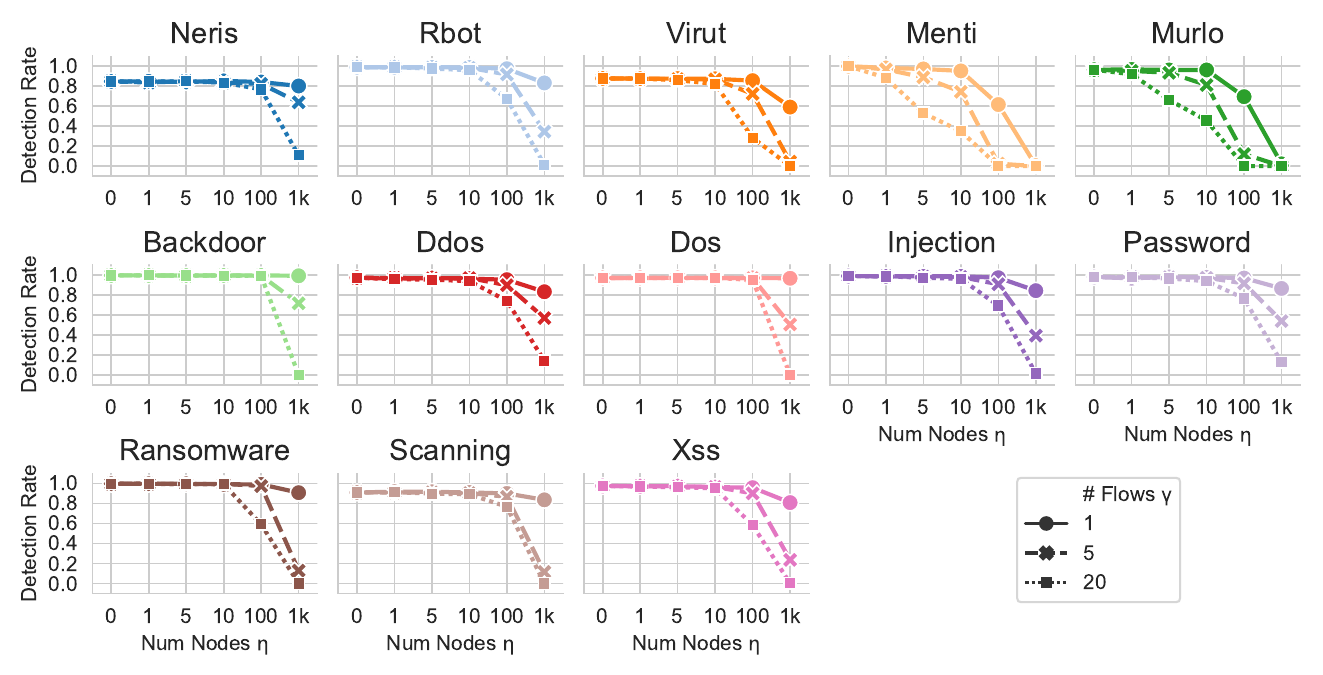}
        \caption{Add node attack against the E-GraphSAGE model.}
        \label{fig:esage_add_node}
    \end{subfigure}

    \begin{subfigure}[htbp]{\textwidth}
        \centering
        \includegraphics[width=1.3\textwidth, center]{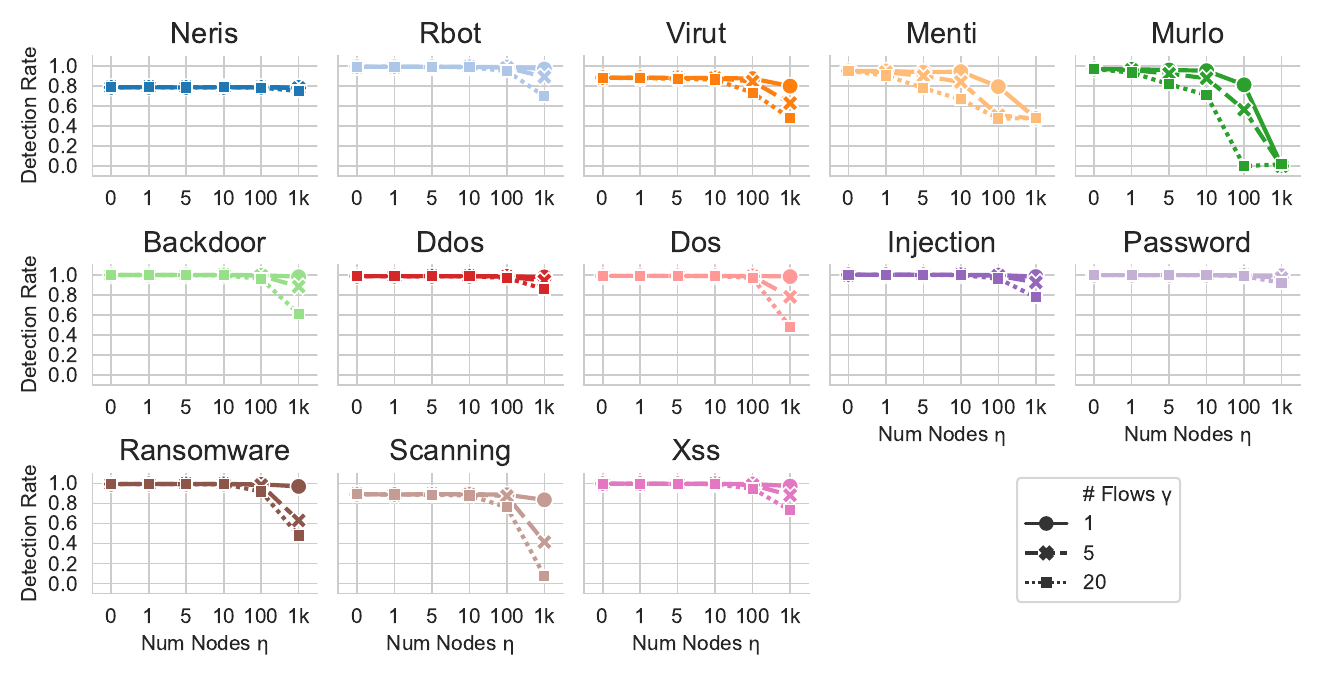}
        \caption{Add node attack against the LineGraphSAGE model.}
        \label{fig:linsage_add_node}
    \end{subfigure}

    \caption{Results for the add node attack.}
    \label{fig:add_node_results}
\end{figure*}

The results for the attack are reported in Figure~\ref{fig:add_node_results}. In particular, Figure~\ref{fig:esage_add_node} and Figure~\ref{fig:linsage_add_node} report the plots for the E-GraphSAGE and LineGraphSAGE models, respectively.
Each figure is structured similarly to previous ones, with the DR plotted on the y-axis and the count of $\eta$ new nodes on the x-axis. The value of $\eta = 0$ indicates the original graph without any attack interference. Each graph shows three distinct lines representing different values of the $\gamma$ parameter.

From the plots in Figure~\ref{fig:esage_add_node}, it is apparent that the Add Node attack tends to be less efficient against E-GraphSAGE with respect to \CtoXB~and \CtoXM~attacks. While the insertion of a single edge in the flow graph contributed to a large performance deterioration, the DR seems unaffected by the addition of a single new node. Nonetheless, as the $\eta$ parameter grows, the drop in the DR becomes more consistent. This is evident especially when $\eta \ge 100$ and $\gamma \ge 5$. Notably, when $\eta=1000$ and $\gamma = 20$, the DR approaches $0$ in almost all the attacks considered.
The LineGraphSAGE model confirms its superior resilience to structural attacks in this scenario as well. As shown in Figure~\ref{fig:linsage_add_node}, even for large values of $\eta$ and $\gamma$, only a few classifiers experience significant performance degradation. The most vulnerable classifiers are those designed to detect the Murlo botnet within the CTU dataset, and the Scanning attack in the ToN-IoT dataset, where we can observe sharp declines in DR.

\subsection{Discussion}
Overall, we can conclude that the considered classifiers are vulnerable to our proposed structural adversarial attacks. 
Our findings yield several crucial insights. First, the Adding Edge attacks, and especially the \CtoXB~variant, are highly effective against the considered classifiers. In particular, we show how it is sufficient for an attacker to set up a negligible number of new communications with random nodes to severely impact the ability of the classifiers to detect malicious samples. Conversely, Adding Node attacks tend to necessitate a large number of newly injected nodes to induce significant decreases in DR.   

We also observe that the two considered models offer different resilience. NIDS based on the LineGraphSAGE algorithms exhibit the highest robustness. However, this enhanced performance comes at the price of the linearization process required to construct line graphs, which can be computationally expensive and can generate very large graphs. Nonetheless, also these classifiers are shown to be vulnerable to the considered attacks, particularly when the perturbations are moderately large. On the other hand, the E-GraphSAGE models are more cost-effective, but they exhibit a marked sensitivity to the graph perturbations. This may be due to the fact that flow graphs provide a natural representation of network traffic. Hence, the topological patterns of cyberattacks are directly influenced by our perturbations. Conversely, we deduce that the linearization procedure results in a representation that is less susceptible to certain types of perturbations.

Finally, in Table~\ref{tab:comparison}, we present a comparison of the results obtained by our adversarial attacks with those presented in related papers. A comprehensive review of the literature reveals only three previous studies that consider adversarial attacks against GNN-based IDS (i.e.,~\cite{zhou2021hierarchical},~\cite{pujol2022unveiling} and~\cite{wang2022threatrace}). These works will be discussed in greater detail in the related work section (Section~\ref{sec:related}). Our comparison in Table~\ref{tab:comparison} includes the type of perturbation (i.e., structural or feature-based), the verification of problem space constraints, the highest metrics achieved in non-adversarial baseline contexts, and the most significant performance drop post-attack. Our investigation highlights that all the previous studies limit their evaluation on feature-based adversarial attacks. Notably, only the work in~\cite{pujol2022unveiling} takes into account feature modifications that do not compromise the malicious functionalities of the adversarial samples. Moreover, we observe that only the attacks in~\cite{zhou2021hierarchical} have a significant impact on the models, with Precision scores going from $0.85$ to $0.5$ in the most severe cases. In contrast, our structural attacks demonstrate the largest performance degradation among the papers reviewed, potentially reducing the DR to zero even with minimal perturbations.

\begin{table}[htbp]
    \centering
    \caption{Performance comparison for adversarial attacks against GNN-based NIDS.}
    \label{tab:comparison}
    \resizebox{0.99\linewidth}{!}{
    \begin{tabular}{|c|c|c|c|c|}
        \hline
        \rowcolor[HTML]{EFEFEF} 
        \textbf{Framework} & \textbf{Perturbation} & \textbf{Constraints verified?} & \textbf{Baseline} & \textbf{Adversarial} \\ \hline \hline
        \cite{zhou2021hierarchical} & Feat. & \xmark & Prec = $0.85$ & Prec = $0.5$ \\ \hline
        \cite{pujol2022unveiling} & Feat. & \checkmark & W-F1 = $0.99$ & W-F1 = $0.99$ \\ \hline
        \cite{wang2022threatrace} & Feat. & unknown & FNR = $0.04$ & FNR = $0.07$ \\ \hline \hline
        \textbf{OURS} & Struct. & \checkmark & F1 = $0.95$ & DR = $0$ \\ \hline
    \end{tabular}
    }
\end{table}

\section{Related Work}
\label{sec:related}

Structural attacks against GNN models are not entirely new~\cite{sun2022adversarial}. Numerous works have proposed and successfully launched such types of attacks against GNN in other domains~\cite{chang2020restricted, zhao2021structural, li2023black, zhang2022semantics}. Nevertheless, their application in the network intrusion domain has not been explored before. Moreover, this domain poses additional constraints that must be addressed in order to devise realistic solutions. In this paper, we identify and discuss these constraints, and we are the first to propose structural attacks that are feasible in practice.   

After a thorough revision of related works, we observe that only a limited number of papers have proposed or considered adversarial attacks against GNN-based (N)IDS (i.e.,~\cite{zhou2021hierarchical, pujol2022unveiling, wang2022threatrace}). Moreover, all of them focus exclusively on feature attacks. The work in~\cite{zhou2021hierarchical} proposes hierarchical adversarial attacks against GNN-based NIDS with a particular focus on the IoT domain. The methodology requires an attacker able to construct a shadow GNN model from intercepted network packets. Once the surrogate model is obtained, the attacker generates adversarial examples capable of fooling it, and submits them to the original target detector, exploiting the transferability property of adversarial attacks~\cite{papernot2017practical}. Conversely, our structural attacks do not require the additional burden of building a shadow GNN model. Moreover, the authors propose to select nodes with high hierarchical values to be the target of the attacks, which may not be feasible in real-world scenarios. In contrast, our approach involves choosing already compromised nodes, offering a more practical and realistic solution. Finally, their work is limited to feature perturbations, while our study encompasses both feature and structural modifications. 
The work in~\cite{pujol2022unveiling} considers a feature-based attack against the GNN-based NIDS proposed in the same paper, aiming to demonstrate its resilience in adversarial environments. Although the primary goal of the paper is not the development of adversarial attacks, the results prove the robustness of GNN models against such attacks. Similarly, in~\cite{wang2022threatrace}, the aim is not to propose adversarial attacks but to introduce a novel approach based on GraphSAGE for host intrusion detection systems (HIDS). Despite this, the evaluation of the model's robustness takes into account also feature-based adversarial attacks. The results align with those of~\cite{pujol2022unveiling}, highlighting the resilience of the GNN.
\section{Conclusions}
\label{sec:conclusion}
Previous research has established the effectiveness of GNN-based NIDS in non-adversarial contexts, with a few studies extending their testing to feature-based attacks.
In this paper, we introduce a new type of adversarial threat, presenting the first formalization of structural attacks specifically tailored for GNN-based NIDS. We identify and discuss the problem-space constraints that come into play when devising realistic structural adversarial attacks in practice. Moreover, we design four structural attack variants that conform to the practical limitations posed by the network intrusion detection domain.
We implement and launch the discussed attacks against two inductive GNN-based NIDS that reach top-level detection performance in two publicly available datasets. The results confirm the enhanced robustness of these models when considering traditional feature-based attacks, which have been proven effective against standard ML-based NIDS. At the same time, however, we show that these models are highly vulnerable to structural perturbations. The application of the proposed structural attacks leads to a severe drop in the detection rates of malicious samples, even with minimal modifications. We demonstrate that our attacks are more effective than those previously proposed by related literature. Our findings uncover a critical vulnerability when taking into account graph representations in the detection processes of ML and DL models, paving the way for future improvements of these techniques. 

\bibliographystyle{abbrv}
\bibliography{references}

\end{document}